\documentclass[12pt]{article}
\usepackage{graphicx}
\usepackage{bm}        
\usepackage{amssymb}   
\usepackage{color}

\def\beq{\begin{eqnarray}}
\def\eeq{\end{eqnarray}}

\def\ba{\begin{eqnarray}}
\def\ea{\end{eqnarray}}
\def\beq{\begin{eqnarray}}
\def\eeq{\end{eqnarray}}

\def\mpl{M_{\rm P}}

\def\p{{\cal P}}

\def\L*{{\cal L}_*}
\def\L{\mathcal{L}}
\def\({\left(}
\def\){\right)}

\def\nn{\nonumber}
\def\p{\partial}
\def\mn{_{\mu \nu}}
\def\stu{St\"uckelberg }
\def\p{\partial}

\def\<{\langle}
\def\>{\rangle}

\newcommand{\be}{\begin{equation}}
\newcommand{\ee}{\end{equation}}

\newcommand{\bea}{\begin{eqnarray}}
\newcommand{\eea}{\end{eqnarray}}
\newcommand{\beas}{\begin{eqnarray*}}
\newcommand{\eeas}{\end{eqnarray*}}

\def\({\left(}
\def\){\right)}

\newcommand{\half}{\frac{1}{2}}

\newcommand{\xb}{\mathbf{x}}

\def\lsim{\mathrel{\rlap{\lower3pt\hbox{\hskip0pt$\sim$}}
     \raise1pt\hbox{$<$}}}         
\def\gsim{\mathrel{\rlap{\lower4pt\hbox{\hskip1pt$\sim$}}
     \raise1pt\hbox{$>$}}}         

\def\lsim{\mathrel{\rlap{\lower3pt\hbox{\hskip0pt$\sim$}}
     \raise1pt\hbox{$<$}}}         
\def\gsim{\mathrel{\rlap{\lower4pt\hbox{\hskip1pt$\sim$}}
     \raise1pt\hbox{$>$}}}         

\usepackage{amsmath}
\usepackage{amsfonts}
\usepackage{verbatim}
\usepackage{setspace}
\singlespace
\begin{document}

\begin{titlepage}

\begin{flushright}
{NYU-TH-09/18/11 \\ UCSD-PTH-12-03}

\end{flushright}
\vskip 0.9cm

\centerline{\Large \bf Classical Duals of Derivatively Self-Coupled Theories}
\vskip 0.7cm
\centerline{\large Gregory Gabadadze$^{a,b}$, Kurt Hinterbichler$^c$, David Pirtskhalava$^d$} 
\vspace{0.1in}
\vskip 0.3cm

\centerline{\em $^a$Center for Cosmology and Particle Physics,
Department of Physics,}
\centerline{\em New York University, New York,
NY, 10003, USA}

\centerline{\em $^b$ Simons Center for Geometry and Physics,}
\centerline{\em Stony Brook University, Stony Brook,
NY, 11794, USA}

\centerline{\em $^c$Center for Particle Cosmology, Department of Physics and Astronomy,}
\centerline{\em University of Pennsylvania, Philadelphia, Pennsylvania 19104, USA }

\centerline{\em $^d$Department of Physics, University of California, San Diego, La Jolla, CA 92093 USA}

\centerline{}
\centerline{}

\vskip 1.cm

\begin{abstract}

Solutions to scalar 
theories with derivative self-couplings often have regions where non-linearities are important.
Given a classical source, there is usually a region, demarcated by the Vainshtein radius, 
inside of which the classical non-linearities are dominant, while quantum effects are still negligible.  
If perturbation theory is used to find such solutions, 
the expansion generally breaks down as the Vainshtein radius is approached from the outside.
Here we show that it is possible, by integrating in certain auxiliary fields, to reformulate these 
theories in such a way that non-linearities become small inside the Vainshtein radius,  and large outside it.  
This provides a complementary, or classically dual, description of the same theory -- one in which 
non-perturbative regions become accessible perturbatively. 
We consider a few examples of classical solutions with various symmetries, and find that in all the  
cases the dual formulation makes it rather simple to study regimes in which the original perturbation theory 
fails to work. As an illustration, we reproduce by perturbative calculations some of the already known 
non-perturbative results, for a point-like  source, cosmic string,  and domain wall,  and derive a new one.
The dual formulation may be useful for  developing the PPN formalism in the theories
of modified gravity that give rise to such scalar theories.

\end{abstract}


\end{titlepage}

\newpage

\section{Introduction and summary}

Perturbation theory is often the only analytic tool available to extract detailed information 
from interacting theories. The regime in which perturbation theory is valid is usually  
limited.  In certain cases, however,  it is  possible to reformulate a theory
in terms of new, dual, variables that allow  perturbative calculations 
in the regime where the original formulation  was non-perturbative.

In this note, we discuss certain special nonlinear theories, and show that  at the classical level 
they admit a dual description. These are  field theories of a scalar, $\phi$, with purely derivative  
nonlinear terms, that nevertheless give equations of motion with no more than two time
derivatives.

Our main motivation for considering such  models, and their classical duals,  
stems from the theories that modify General Relativity (GR) in the infrared   --  the five-dimensional 
DGP model \cite{Dvali:2000hr},  and four-dimensional ghost-free massive gravity 
\cite{deRham:2010ik,deRham:2010kj}.
The four-dimensional scalar Lagrangians discussed here capture parts 
of the full gravitational theory,  as shown in \cite{Luty:2003vm} for DGP and \cite{deRham:2010ik} for
massive gravity (for reviews and experimental limits, see Refs. \cite{Goldhaber:1974wg}, 
and for a recent theoretical review of massive gravity see Ref. \cite{Hinterbichler:2011tt}).

Our analysis may have  broader applications though: derivatively self-coupled theories, in particular
the galileons \cite{Luty:2003vm,Nicolis:2004qq,Nicolis:2008in},  can also be obtained 
in the probe-brane limit of higher dimensional constructions \cite{deRham:2010eu}, and  their 
extensions \cite{Hinterbichler:2010xn,VanAcoleyen:2011mj,Goon:2011qf,Goon:2011uw,Trodden:2011xh}, 
while their three-dimensional counterparts are obtained  in 
the context of  three-dimensional ``new massive gravity'' \cite{Bergshoeff:2009hq}, as 
shown in \cite{deRham:2011ca}.
 
\vspace{0.1in}

In the present  work we shall focus on the so-called cubic galileon,  
$\sim\square\phi(\partial\phi)^2/\Lambda^3,$ \cite{Luty:2003vm,Nicolis:2004qq} 
by which the  free Lagrangian, $ -(\partial \phi)^2/2 $, is supplemented
in four-dimensions.  The state  described by $\phi$ can be thought of as a Nambu-Goldstone boson, 
nonlinearly realizing (a limit of) broken higher dimensional  Poincar\'e or diffeomorphism invariance 
$$\phi\to\phi+c+c_{\mu} x^{\mu},$$ with $c_{\mu}$ denoting a constant vector. 
In parallel, we will also consider -- mainly as toy examples -- 
theories of an ``ordinary" Nambu-Goldstone (NG) 
boson with the self-interaction terms, such as $- (\partial\phi)^{2n}/\Lambda^{4n-4},~n\geq 2$. 

The only physical scale in these models is $\Lambda$. Such theories are usually  regarded 
as effective field theories valid at energies/momenta below $\Lambda$. One reason for this is that scatterings 
of the $\phi$-quanta --   when treated in  conventional perturbation theory --  exhibit non-perturbative 
behavior at/above the scale $\Lambda$. On the other hand, the galileons 
do not seem to represent garden variety effective field theories.  They 
are special -- for instance, they do not  get renormalized by quantum corrections
\cite{Luty:2003vm,Nicolis:2004qq,Nicolis:2008in,Hinterbichler:2010xn} 
(although, other higher-derivative terms may be generated). 
One  may wonder then, if  there may be some hidden structure
in the galileon  theories that would enable one to deal in a controllable way  
with scales above $\Lambda$,  by a re-summation of perturbative diagrams,  
or by a dual description.  Although, in the present work, we will not 
explore the above important question, we'll make a step in that direction.

What we shall show, instead, is that a  dual description is possible for these theories 
in the classical regime. To enable a window in which  the classical description is meaningful,  
we introduce a minimal coupling of the scalar $\phi$ to the trace of an external classical stress-tensor 
$T$ (planets, stars,  etc.)
\beq
\sim{ \phi \over M_{\rm Pl}} T \,,
\eeq
where $M_{\rm Pl}$ is the Planck  mass. Coupling to the stress-tensor can be non-minimal in a  
more general context \cite{deRham:2010ik}, with  certain interesting observational 
consequences \cite{Wyman:2011mp}; 
our analysis should straightforwardly apply to those cases too.

The presence of the Planck mass, in addition to $\Lambda$, gives rise to a new {\it derived} scale,  
referred to as the Vainshtein scale \cite{Vainshtein:1972sx}. For a static spherically symmetric 
classical source of mass $M$,  $r_\ast \sim  \Lambda^{-1} (M/ M_{\rm Pl})^{1/3}$ \cite{Deffayet:2001uk}. 
This scale is much greater than $\Lambda^{-1}$. The conventional perturbative 
expansion can be used to compute the field configuration  outside the Vainshtein radius,  $r>> r_\ast$. 
Inside the radius, $r<<r_\ast$,  classical non-linearities in $\phi$  are  dominant, and 
the perturbative expansion breaks down. More formally speaking, external classical 
sources introduce a new expansion parameter, $\alpha_{cl}$, that captures the  strength 
of classical nonlinearities;  for the galileon, $\alpha_{cl}=\p^2\phi/\Lambda^3$,
while for the NG-type  theories, $\alpha_{cl}=(\p\phi)^2/\Lambda^{4}$.  
The parameter  $\alpha_{cl}$ is source-dependent, and for theories considered here, 
there is generically a broad region in space 
where $\alpha_{cl}\geq 1$, while energies and momenta are still well-below $\Lambda$. 
Therefore, the classical field enters a highly nonlinear regime, while the  quantum 
corrections are still negligible, as long as  we stay at 
the distance scales  greater than $\Lambda^{-1}$. 

We show how these theories can be dualized by integrating in certain auxiliary variables. 
The dual  theory is classically equivalent to the original one, however,  it no longer has 
any higher-dimensional  derivatively-coupled terms. Instead, the dual theory is nonlocal, 
in a sense that it contains lower dimensional non-derivative 
terms with fractional classical dimensions. 

Perturbation theory in the dual version has a regime of validity opposite to the original one: 
there is still a Vainshtein radius, but now non-linearities are small \textit{inside} the Vainshtein radius, 
and large outside of it.  Hence, the non-perturbative regime in the original variables is perturbative 
in the dual picture, and {\it vice~versa}.

We point out that, whether in the dual description or the original description, in both the perturbative 
and non-perturbative regimes the classical fields are {\it weak} in Planckian units, $\phi_{cl}<< M_{\rm Pl}$.
This should be compared to the case of a  black hole in GR, where nonlinearities 
near the Schwarzschild radius, $r_g$,  are due to classical fields that aren't  small 
in Planckian units.  These nonlinearities can be re-summed  into 
the Schwarzschild solution\footnote{It may be interesting to attempt  
to dualize  GR along the lines discussed here.}.

\vspace{0.1in}
 
We would like to make a few important comments on the literature.
First, our dual description  is what captures the properties of the  small-mass  
expansion used by Vainshtein \cite{Vainshtein:1972sx} in massive gravity. The latter expansion 
is replaced here by a series governed by positive powers of $\Lambda$. This  parameter,  in the context of 
massive gravity, is derived from the graviton mass and Planck scale.  Second,
the  derivatively coupled theory of a $\phi$ field containing a nonlinear 
ghost (for instance the Lagrangian $\sim(\square \phi)^3/\Lambda_5^5$ describing the longitudinal 
mode of earlier, ghostly massive gravity theories), was decomposed  by Deffayet and Rombouts  
\cite{Deffayet:2005ys} by means of the Ostrogradskii method to manifestly exhibit the ghost in 
the linear theory.  Our construction is similar, but not identical,  since we're dealing with the 
theories without ghosts; this essential distinction gives rise 
to significant differences in the two  cases, as will be seen below. In spite of the differences, 
we have been inspired by both \cite{Vainshtein:1972sx} and \cite{Deffayet:2005ys}.

The work is organized as follows: In Section 2 we consider the simplest NG-type model.
We discuss its classical  dual and perform calculations for  sources with spherical and cylindrical 
symmetry. This serves demonstrational purposes, as the main focus of the present work is on galileons
which are relevant to the theories of IR modified gravity. In Section 3 we study the cubic galileon. 
Again, we present the dual theory, and use it for perturbative calculations with spherical, 
cylindrical,  and planar symmetries. Section 4 contains  conclusions and outlook. In the Appendix
we discuss more general NG-type theories.

\section{Nambu-Goldstone type theories}

As mentioned before, the galileons \cite{Luty:2003vm,Nicolis:2008in},  are perhaps the most remarkable 
derivatively self-coupled scalar field theories:  their  special structure 
guarantees  a good Cauchy formulation, as well as 
non-renormalization of these terms \cite{Luty:2003vm,Nicolis:2008in} in the quantum theory with no 
additional fields. 
Quantum effects  may generate other derivative terms,  such as $(\square \phi)^k,~~k \geq 2$, however, 
the effects of the latter are suppressed in the classical regime  considered here. Therefore, our restriction 
to a single cubic galileon term in the next section can be justified even in the full quantum theory.

On the other hand, there is no  similar argument  for the NG-like theories. 
Also, there is no known principle which would lead to the re-summation of the $(\p\phi)^{2n}$ -- type 
NG interactions, except in the case leading to the  DBI action.  DBI, however,  
has no well-behaved static solutions \cite{Goon:2010xh},  and we will not consider it here.
Therefore, restricting  only to the NG-type term with $n=2$, as it will be done in this  section,  
is not  justified in the full quantum theory.
Nevertheless, we consider this example as a starting point in this section  
and regard it as  a toy model where calculations are easier, 
keeping in mind that generically one should  be retaining terms with all 
possible integer values for $n$ (other values of  $n$ are considered in the Appendix).

Thus,  we consider  a theory of a scalar field $\phi$ with a NG-type derivative quartic 
self-interaction
\beq
\L_1=-\frac{1}{2•}(\p\phi)^2-\frac{1}{4 \Lambda^{4}}(\p\phi)^4+\frac{1}{\mpl}\phi T.
\label{goldstone}
\eeq
Around the trivial background, $T=0,~\phi=0$, perturbation theory for the amplitudes 
of the $\phi$-quanta starts to fail as energies reach the scale $\Lambda$. We now turn to 
nontrivial classical backgrounds with nonzero $T$~\footnote{Note that we consider this model  
with a ``wrong sign'' in front of the nonlinear term -- the sign that does not admit a 
conventional UV completion \cite{Adams:2006sv}, while it exhibits the Vainshtein mechanism 
\cite{gia1}.  We thank Lasha Berezhiani for pointing out that 
the ``right-sign'' nonlinear term  for the NG,  $ + (\partial \phi)^4/\Lambda^4$,  
does not admit  the Vainshtein mechanism.

The theory (\ref {NG1}) with $T=0$ can be regarded as the decoupling 
limit of a massive abelian vector field with a quartic interaction \cite{gia2},
\beq
\mathcal{L}=-\frac{1}{4} F^{\mn} F_{\mn}-\frac{1•}{2•} m_A^2 A^{\mu}A_{\mu}-\frac{g^4}{4•}\( A^{\mu} A_{\mu} \)^2.
\label{NG1}
\eeq
At energies parametrically above $m_A$, $\phi$ describes the helicity-0 
component of $A_\mu$, extracted through the St\"ukelberg replacement,
$
A_{\mu}\rightarrow \tilde{A_\mu}-\frac{1}{m_A•}\p_\mu\phi,
$
and the decoupling limit is defined as follows,
$
 \quad m_A \to 0,\quad g\to 0, \quad \Lambda \equiv  \frac{m_A}{g•}\ {\rm fixed}.
$
The effective theory \eqref{goldstone} is valid at distance scales $\Lambda^{-1}\ll r \ll m_A^{-1}$.}.

The equation of motion reads
\be \square\phi+\frac{1}{ \Lambda^{4}}\partial_\mu\left[(\p\phi)^2\partial^\mu\phi\right]
=-{T\over \mpl}.\label{fulleqgold}\ee
For a static point-like  source, $T=-M\delta^3(\bf{ x})$, the solution for $\phi$ is spherically symmetric 
and static, and the equation of motion reduces to 
\beq
\vec\nabla\cdot\(\vec\nabla\phi+\frac{1}{\Lambda^4•}
(\vec\nabla\phi)^2\vec\nabla\phi\)=\frac{M}{\mpl•}\delta^3(\bf{ x}),
\label{eqn0}
\eeq
which can readily be integrated once to obtain a cubic algebraic equation for the radial derivative $\phi'$,
\beq\label{radialalg}
\phi'+\frac{1}{\Lambda^4•}(\phi')^3=\frac{M}{4\pi\mpl}\frac{1}{r^2•}.
\eeq 
This can be solved exactly. The exact solution has two regimes, depending on which of 
the two terms on the left hand side of (\ref{radialalg}) dominates.  
The scale that separates the two regimes is denoted $r_*$, this 
being the distance at which the two terms become comparable,
\beq
\label{vainsteinrgold}
r_*\sim \(\frac{M}{\mpl}\)^{1/2}\frac{1}{•\Lambda}.
\eeq
At scales larger than $r_*$, the linear term on the l.h.s. dominates, leading to the usual Newtonian potential 
for the scalar 
\beq
\phi\simeq -\frac{M}{4\pi\mpl}\frac{1}{r•}, \qquad r\gg r_*.
\eeq
Note that the value of the  classical field is small in Planckian units: ${\phi \over  M_{\rm Pl}}\sim 
{r_g \over r} << 1$,  for $r >  r_\ast$.
At distances shorter than $r_\ast$ on the other hand, the non-linear term is more important, 
and the solution  reads
\beq
\phi\simeq 3  \(\frac{M}{4\pi\mpl}\)^{1/3} \Lambda^{4/3} r^{1/3}+const.,   \qquad r\ll r_*.
\label{0}
\eeq
It is straightforward to check that even in this regime  the value of the classical field 
(\ref {0}) is sub-planckian, ${\Phi \over M_{\rm Pl}} \sim {r_g \over r_\ast} ({r \over r_\ast})^{1/3} <<1$, and 
decreases with decreasing  $r$.

If the exact solution were not known, and we wanted to set up a perturbation theory to find it, we could  
perform an expansion in powers of the interaction.  We could do this by expanding the field into 
powers of the non-linear interaction,
\be \phi=\phi_0+\phi_1+\phi_2+\cdots.\ee
Plugging this expansion into the equation of motion (\ref{fulleqgold}) and equating 
powers of ${1\over \Lambda}$, one generates a series of equations
\bea && \square\phi_0=-{T\over \mpl}, \\ 
&&  \square\phi_1+{1\over \Lambda^4}\partial_\mu\left[(\p\phi_0)^2\partial^\mu\phi_0\right]=0, \\ 
 && \vdots
 \eea
For the static point-like  source  $T=-M\delta^3(\bf{ x})$,  
the leading  order solution is $\phi_0= -\frac{M}{4\pi\mpl}\frac{1}{r•}$.
By simple power counting one can see  that the series is nothing but  an 
expansion in powers of the parameter \be \left( {r_\ast\over r}\right)^4,\label{origparam}\ee
where $r_\ast$ is the Vainshtein radius (\ref{vainsteinrgold}).  
The expansion is good for large radii, and starts to fail 
as we approach the Vainshtein radius from the outside.

\subsection{Dual formulation}

The theory in the form given above naturally yields a perturbative  expansion which is 
valid in the IR but breaks down in the UV.  As discussed above, the region of transition between 
the classically perturbative and non-perturbative regimes lies around the Vainshtein radius
\footnote{We mention again that quantum-mechanically, we should really be considering all operators of the 
form $\(\p\phi\)^{2n}$, which become of the same order as $\(\p\phi\)^2$ once the Vainshtein radius 
is approached from the outside. Since we are concerned here with a mere illustration of how dual theories work 
(before a much more stable analysis of galileons in the next section), and for simplicity, we choose to 
ignore these operators.}.

We would like to rewrite the theory in a form that makes it possible to perform a perturbative expansion 
which is valid in the UV, rather than in the IR.  It is straightforward to check that 
the following Lagrangian written in terms of the original scalar $\phi$,  and an auxiliary vector 
field  $\psi_\mu$,
\beq
\L=-\frac{1}{•2}(\p\phi)^2+\frac{3}{4•}\Lambda^{4/3}(\psi_\mu^2)^{2/3}-\psi^\mu\p_\mu\phi+\frac{1}{\mpl•}\phi T,
\label{dualgold}
\eeq
recovers back the original theory (\ref {goldstone}) upon integrating out $\psi_\mu$.

By simply introducing a new  variable, $\psi_\mu$, we seem to have arrived at an equivalent 
action in which none of the coupling constants (never mind Planck's mass) are of negative mass 
dimension, naively pointing towards the possibility of a perturbative expansion which is good 
in the UV.  Note however that there is now a potential term for $\psi_\mu$ with a fractional power, 
and that this fraction is less than two
(in Appendix \ref{nappendix}, we show that more general interactions also lead to such terms 
and that the fractional power is always less than two).

In the  limit $\Lambda \rightarrow 0,~~M_{\rm Pl} \rightarrow \infty$, the theory (\ref {dualgold})
reduces to the one governed by the non-dynamical equations, 
$\partial_\mu \phi =0,~~\partial^\mu \psi_\mu =0$.

Note that the second term in (\ref {dualgold}), if regarded as a potential, 
has a ``tachyonic'' sign.  This is a consequence of the minus sign in  front of the 
nonlinear term in (\ref {goldstone}), and is a cause of the existence of 
superluminal modes \cite{Adams:2006sv,Nicolis:2008in,Andrews:2010km,Goon:2010xh} in the theory (\ref {goldstone}). Thus, in the dual version 
the superluminality is related to the tachyonic instability  of the non-analytic potential.
The latter could be stabilized, e.g., by supplementing (\ref {dualgold}) with 
carefully chosen higher powers of  $\psi_\mu^2$. 
However, we  will not pursue this completion here,  since  we just use the theory 
(\ref {goldstone}) as a toy example to demonstrate the trick of dualization in an easier 
setup. 

The equations of motion that follow from (\ref {dualgold}) are
\ba
\Box\phi+\p_\mu\psi^\mu=-\frac{1}{\mpl•} T, \nn \\ \Lambda^{4/3} \((\psi_\nu)^2\)^{-1/3}\psi_\mu-\p_\mu\phi=0.
\label{eqs}
\ea
At this point we choose to make the field  decomposition
\beq
\psi_0=\chi, \qquad \psi_i=\psi_i^T+\p_i\psi , \qquad \sigma=\phi+\psi,
\label{3+1}
\eeq
under which the equations become
\ba
\Box\sigma+\p_0^2\psi-\p_0\chi&=&-\frac{1}{\mpl•} T, \nn \\ \Lambda^{4/3} 
\(-\chi^2+(\psi^T_i+\p_i\psi)^2\)^{-1/3}\chi+\p_0\psi&=&\partial_0\sigma, \nn \\ \Lambda^{4/3}
\(-\chi^2+(\psi^T_i+\p_i\psi)^2\)^{-1/3}(\psi^T_i+\p_i\psi)+\p_i\psi&=&\partial_i\sigma.
\label{eqs}
\ea
One can see that the vanishing transverse component of $\psi$, $\psi^T_i=0$, is a consistent ansatz for 
the solution. Moreover,  for static field configurations, $\chi=0$ and 
$\sigma$ obeys a linear equation sourced by $T$, while $\sigma$ in  
turn  determines $\psi$. Instead of the irrelevant operator of the original theory (\ref{goldstone}), 
we have only the self-interaction of the field $\psi$ which looks like a \textit{relevant} operator, 
and we could expect it to be subdominant in the UV.  In the IR, on the other hand, 
things are ill-defined  because the interactions above the trivial ground state are 
non-analytic\footnote{All the above statements refer to the classical  theory.  We point out, 
however, that non-covariant decomposition (\ref {3+1}), which does not introduce extra time derivatives, 
is likely to be a good starting point for quantization of this theory.}.

For the static point source  $T=-M\delta^3(\xb)$,  the only excited degrees of freedom are 
$\sigma$ and the longitudinal component of $\psi_i$.  The first equation in  \eqref{eqs} tells us 
that the exact value of $\sigma$ is its linear Newtonian value, $\sigma= -\frac{M}{4\pi\mpl}\frac{1}{r•}$,  
while the equation for $\psi$  reduces to the following  one
\ba
\psi'+\Lambda^{4/3}(\psi')^{1/3}=\frac{M}{4\pi\mpl}\frac{1}{•r^2}.
\label{eqns2}
\ea  
If we ignore the non-linear interaction by setting $\Lambda=0$,  $\psi$ will have the zeroth order linear 
solution, $\psi_0=-\frac{M}{4\pi\mpl}\frac{1}{r•}$.  The full solution for the original field $\phi_0$ is 
then trivial, since there is a cancellation $\phi_0=\sigma-\psi_0=0$.  It is easy now to estimate the distance 
for which the non-linear term in (\ref{eqns2}) becomes important.   We find that this scale  is again the 
Vainshtein radius 
$r_\ast$ (\ref{vainsteinrgold}). However,  in contrast to the original formulation of the theory, 
the self-interactions are now important only at distances larger than the Vainshtein radius, 
$r\gtrsim r_*$.  

We can see this reversal in the region of strong coupling more explicitly by solving the $\psi$ 
equation perturbatively using the dual formulation.  We set up the expansion by expanding in powers 
of the interaction coupling $\Lambda^{4/3}$,
\be \psi=\psi_0+\psi_1+\psi_2+\cdots,\ee
plugging into (\ref{eqns2}) and equating powers of $\Lambda$.  The solution to lowest non-trivial order is
\beq
\psi_0+\psi_1=-\frac{M}{4\pi\mpl}\frac{1}{r•}-3 \Lambda^{4/3}\(\frac{M}{4\pi\mpl}\)^{1/3}r^{1/3}+const.,
\eeq
while $\sigma$ has the linear Newtonian $1/r$ profile to  all orders. 
Recalling the definition of the physical field $\phi$, we have,
\beq
\phi=\sigma-\psi=3 \Lambda^{4/3}\(\frac{M}{4\pi\mpl}\)^{1/3}r^{1/3}+const.+\cdots~,
\eeq
this shows that the expansion is in powers 
\be \left(r\over r_\ast\right)^{4/3}.\ee
This expansion is inverse to the expansions of the original theory (\ref{origparam}). 
As a result,  the dual expansion  breaks down as we approach the Vainshtein radius from 
the inside.

\subsection{Profile of an infinite string}

The use of the  dual formulation is not restricted to spherically symmetric static solutions.  
It should provide a complementary description for any classical solution.  
For illustrative purposes, we now consider a cylindrically symmetric solution sourced by a uniform string, 
with the mass-per-unit-length  denoted by $\kappa$
\beq
T=-\kappa\frac{\delta(r)}{2\pi r•}.
\eeq
The exact solution again has two regimes, separated by a Vainshtein radius
\beq
r_*\sim \frac{\kappa}{\mpl \Lambda^2•}.
\eeq
The leading behavior of $\phi$ in the two regimes is,
\beq
\phi= \begin{cases} \frac{3}{2•}\(\frac{\kappa}{2 \pi\mpl}\)^{1/3}\Lambda^{4/3} r^{2/3}+const., & r\ll r_\ast ,\\
\frac{\lambda}{2 \pi\mpl} \ln \(\frac{r}{r_s}\), & r\gg r_\ast,
\end{cases}
\label{exact}
\eeq
where $r_s$ is a UV regulator scale -- the transverse width  of the string in this case. 
Using perturbation theory in the original formulation, we recover the logarithmic profile for $r\gg r_\ast$ as the 
leading term, 
and perturbation theory breaks down as we approach the Vainshtein radius from the outside.

On the other hand, the dual form of the equations of motion, \eqref{eqs}, yields the following expressions for the 
fields $\sigma$ and $\psi$,
\begin{eqnarray}
&& \chi=0, \quad \sigma = \frac{\lambda}{2 \pi\mpl} \ln \(\frac{r}{r_s}\), \quad \psi\simeq  
\frac{\lambda}{2 \pi\mpl} \ln \(\frac{r}{r_s}\) - \frac{3}{2•}\(\frac{\lambda}{2 \pi\mpl}\)^{1/3}
\Lambda^{4/3} r^{2/3}+const. \nn \\ 
\end{eqnarray}
The expression for $\sigma$ here is exact as above, while the series for $\psi$ breaks down in the IR, 
as the Vainshtein scale $r_*$ is approached from the inside. Recalling the definition  $\phi=\sigma-\psi$, one again 
finds an agreement with the result obtained in the original formulation of \eqref{exact}.

\section{The cubic galileon}

We have illustrated how the simplest model of a single Nambu-Goldstone scalar can be re-written in a form for 
which classical perturbation theory has a region of validity opposite to that of the original formulation.  
The method however is quite general.  The essence of the method is to introduce auxiliary fields in such a way as to 
replace the non-renormalizable derivative interactions with (generally non-analytic)  non-derivative  
terms. We now consider the cubic galileon \cite{Luty:2003vm,Nicolis:2004qq} - an example of a scalar field theory that is 
relevant for modifications of gravity,
\be 
{\cal L}=-\half (\partial\phi)^2-{1\over \Lambda^3}(\partial\phi)^2\square\phi+{1\over M_P}\phi T.
\label{3g}
\ee
As mentioned  before, in the $M_{\rm Pl} \to \infty $ limit,  the galileon term in the above action  
does not get renormalized  in the 
full quantum theory;  also, no higher galileons \cite{Nicolis:2008in} 
will be induced if they're not introduced to begin with. Moreover, no NG-type terms, $(\partial \phi)^n$,
of the previous section  will be generated since  the latter do not respect the ``Galilean symmetry'' 
$\phi\to\phi+c+c_{\mu} x^{\mu}$.

In spite of the presence of higher derivatives, the equations of motion that follow from the above action 
are second-order, leading to a well-defined Cauchy problem,  and the absence of additional ghostly degrees of freedom,
\be \label{piequation} \square\phi-{2\over \Lambda^3}
\left[(\partial_\mu\partial_\nu\phi)^2-(\square\phi)^2\right]=-{T\over  M_P}.\ee

Concentrating again on radial profiles for $\phi$ sourced by a point-like source, $T=-M\delta^3( \xb)$, 
the equation of motion (\ref{piequation}) reduces to the following, 
\be \vec\nabla\cdot\left[\vec\nabla\phi+{1\over \Lambda^3}
\left(2\vec\nabla\phi\, \vec\nabla^2\phi-\vec\nabla(\vec\nabla\phi)^2\right)\right]={M\over  M_P}\delta^3( \xb).\ee
Integrating once, we obtain  a quadratic equation for the radial derivative of the galileon field,
\be {\phi'\over r}+{4\over \Lambda^3}{\phi'^2\over r^2}={M\over  M_P}{1\over 4\pi r^3}.\ee
Like in the model of the previous section, the exact solution has two regimes, 
separated by the Vainshtein radius $r_\ast$,
\be \label{sphericalsol}
\phi(r)= \begin{cases}{1\over 2\sqrt{\pi}} \Lambda^3 r_*^{2} \left(\frac{r}{r_*}\right)^{1/2}\left[1+{\cal O}\(r^{3/2}\over r_*^{3/2}\)\right]+const. & r\ll r_*, \\ -{M\over M_P}{1\over 4\pi r}\left[1+{\cal O}\(r_*^3\over r^3\)\right] & r\gg r_*,\end{cases} \ 
\ee
where
\be r_*\equiv\left(M\over M_{P}\right)^{1/3}{1\over \Lambda}.\label{galvainshtein}\ee
The Vainshtein mechanism is therefore at work in the cubic galileon theory as well, screening 
the scalar potential significantly within $r_*$.  As in the previous section, the classical field 
is  weak (sub-Planckian) in both the linear and non-linear regimes, $(\phi/M_{\rm Pl})<< 1$.

Perturbation theory in this formulation allows us to compute the corrections to the ${1/ r}$ 
solution for $r\gg r_*$ via a $1/ \Lambda^3$ expansion of the equation (\ref{piequation}).  We thus write,
\be \phi= \phi_0+\phi_1+\phi_2+\cdots,\ee
so that after plugging into the equation of motion and equating powers of $1/\Lambda$, we find the series of equations  
\bea \label{piequationpert} && \square\phi_0=-{T\over  M_P}. \\
&& \square\phi_1-{2\over \Lambda^3}\left[(\partial_\mu\partial_\nu\phi_0)^2-(\square\phi_0)^2\right]=0. \\
&&\vdots 
\eea
This gives an expansion in powers of $\(r_*/r\)^{3}$, which is valid outside the Vainshtein radius and starts to fail 
as the Vainshtein radius is approached from the outside.

\subsection{The dual galileon theory}

We would now like to find a dual formulation of the galileon, one whose classical perturbative 
expansion is valid \textit{inside} the Vainshtein radius.  
Starting with the original Lagrangian \eqref{3g}, we introduce two auxiliary scalar fields $b_\mu$ and $\lambda$ 
and write  an equivalent version of the theory as follows,
\be 
\label{lag4} 
{\cal L}=-\half (\partial \phi)^2 +\Lambda^{3/2}\sqrt{\lambda b_\mu^2} 
-b^\mu\partial_\mu\phi - \lambda\square\phi+{1\over M_P}\phi T.
\ee
Again, we have succeeded in representing the cubic galileon in a 
form in which all terms look \textit{relevant} at the expense 
of introducing  fractional dimensions on fields.
This has a more complicated structure than the dual  Lagrangian of the previous section
(\ref {dualgold}). However, there are some similarities, such as that the 
nonanalytic potential term in (\ref {lag4}) also has a tachyonic sign, 
and the fact  the $\phi$ field  becomes trivial in  both (\ref {dualgold}) and (\ref {lag4}) 
in the limit $\Lambda \to 0$,  $M_{\rm Pl} \to \infty$.  This, and the spectrum of the theory, 
is best seen by using a non-covariant (i.e., a 3+1) decomposition of the vector field $b_\mu$, 
as given  below\footnote{As noted in the previous section, this decomposition, unlike the covariant one,  
does not introduce artificially  extra time derivatives which  would  
have confused the counting of propagating modes already in the classical theory.}.  

The equations for $\phi$, $\lambda$ and $b^\mu$ that follow from the latter Lagrangian are given as follows,
\ba
\label{geq}
\Box\phi+\p_\mu b^\mu-\Box\lambda=-\frac{1}{\mpl•} T, \nn \\
-\Box\phi+\frac{1}{2•}\Lambda^{3/2}\sqrt{\frac{b^2_\mu}{\lambda}}=0,  \\
-\p_\mu\phi+\Lambda^{3/2}\sqrt{\frac{\lambda}{b^2_\nu}}b_\mu=0.  \nn
\ea
Resorting again to the $3+1$ decomposition of the vector field, taking the divergence of the last of the equations of motion \eqref{geq}\footnote{For the static profiles at hand, retaining only the divergence of this equation is sufficient for obtaining the complete solution.}, forming suitable linear combinations and (re)defining fields as follows,
\beq
b_0=\beta, \quad b_i=b^T_i+\p_i b, \quad \lambda=\bar \lambda+b, \quad \phi=\bar \phi+\bar \lambda,
\eeq
one can reduce \eqref{geq} to the following system,
\ba \label{syst1}
\square\bar \phi-\p_0\beta+\p_0 b=-\frac{1}{•\mpl} T, \nn \\
\p_0\(\p_0\bar\lambda-\Lambda^{3/2}\sqrt{\frac{\bar\lambda+b}{b^2_\nu•}}\beta\)-\p_i\(\p_i \bar \lambda-\Lambda^{3/2}\sqrt{\frac{\bar \lambda+b}{b_\nu ^2}}\(b^T_i+\p_i b\)\)\nn \\+\p^2_0 b-\p_0\beta=-\frac{1}{\mpl•} T,   \\
\frac{1}{2•}\sqrt{\frac{b_\nu^2}{\bar \lambda+b}}-\p_0\(\sqrt{\frac{\bar\lambda+b}{b^2_\nu}}\beta\)+\p_i\(\sqrt{\frac{\bar \lambda+b}{b_\nu ^2}}\(b^T_i+\p_i b\)\)=0.\nn
\ea
As in the example of the previous section, in the presence of a point source, $T=-M\delta^3(\xb)$, 
the transverse component $b^T_i$ as well as $\beta$ vanish; moreover, one combination of the fields, denoted by 
$\bar \phi$,  is free and has a linear equation everywhere in space, receiving an exact Newtonian profile
\beq
\bar \phi=-\frac{M•}{4\pi \mpl•} \frac{1}{•r}.
\eeq
The last equation from the system \eqref{syst1} on the other hand gives
\beq
b=-\bar\lambda-\frac{1}{4}r\bar \lambda',
\eeq
which, when plugged into the second of \eqref{syst1}, gives an equation for $\bar\lambda$ after integrating once\footnote{One has to be careful with the square root at this point. Jumping a bit ahead, we note that (in complete analogy to the Goldstone-\stu case considered above) the leading term in the expression for $\bar\lambda$ well within the Vainshtein radius is of the form $A/r$, with $A$ denoting some positive constant. This uniquely fixes all signs in front of square roots, as well as makes the $\sqrt{\bar \lambda + b}$ expression in these equations well-defined.},
\beq
-\bar\lambda'+\Lambda^{3/2}\sqrt{-\frac{1}{4•}r\bar\lambda'}=\frac{M}{4\pi\mpl•}\frac{1}{r^2•}.
\label{leq}
\eeq

We have now arrived at an equation which achieves the goal of the dual formulation.  The interaction term, proportional to $\Lambda^{3/2}$, becomes important only at distances \textit{larger} than the Vainshtein radius (\ref{galvainshtein}).
We again set up perturbation theory by expanding the $\bar \lambda$-profile in powers of $\Lambda^{3/2}$,
\beq
\bar\lambda = \bar\lambda_{0}+\bar\lambda_{1}+\bar\lambda_{2}+\dots~.
\eeq
Plugging the expansion into \eqref{leq} and equating powers of $\Lambda$, we obtain
\beq
\bar \lambda_{0}=\frac{M}{4\pi\mpl•}\frac{1}{•r}, \qquad \bar \lambda_{1}=\Lambda^{3/2}\(\frac{M}{4\pi\mpl}\)^{1/2} r^{1/2}+const., \qquad \cdots
\eeq
Finally, recalling the definition of the physical field $\phi$, we have
\beq
\phi=\bar\phi+\bar\lambda=\frac{1}{2\sqrt{\pi}} \Lambda^3 r_*^{2} \left( \frac{r}{r_*} \right)^{1/2}+const.+\cdots,
\eeq
in complete agreement with the result \eqref{sphericalsol} of the original theory well within the Vainshtein radius.  The perturbative expansion in the dual formulation is an expansion in the ratio 
\be \left(r\over r_\ast\right)^{3/2},\ee
and so the expansion is valid inside the Vainshtein radius,  
complementary to the expansion in the original formulation which is valid outside $r_*$.

\subsection{Domain wall and infinite string}

Similarly to the case of the Nambu-Goldstone theory, the dual formulation should be useful in reorganizing the perturbation expansion for any classical solution.  It is interesting to see this on examples other than that of a point source - such as a domain wall or a string. 

It has been shown that domain walls do not possess a Vainshtein scale in DGP and massive gravity \cite{Dvali:2006if}
(this scale is of the order of the wall width), so they give rise to a fifth force at all distances. This fact should be captured by the cubic galileon theory \eqref{3g} and therefore by its dualized version, presented above. The absence of an $r_*$ distance can be easily seen in the equations of motion of the original theory  \eqref{piequation}. Indeed, in the presence of an infinite domain wall in the $x-y$ plane at $z=0$, the problem becomes one - dimensional, with $\phi$ depending on a single coordinate $z$. One can then easily see that all nonlinearities in the original equations of motion vanish and one is left with the Newtonian profile of a 1D source for the scalar. On the other hand, the last of the equations of motion \eqref{syst1} that follow from the dual theory implies $\lambda=0$ for only $z$ - dependent profiles. Recalling the definition of the original galileon $\phi$ in terms of the free scalar $\bar\phi$ of the dual theory and $\lambda$, $\phi=\bar\phi+\lambda$, one obtains agreement between the two representations of the theory (as must be the case since the representations are classically equivalent).  Both the original and dual formulations are free of non-linearities, and so there is no perturbative expansion to be done in either case.

Next consider an infinite string source, $T=-\kappa\frac{\delta(r)}{2\pi r•}$. In the original galileon theory, the equation of motion can be integrated to yield the following algebraic equation for the radial derivative of the axially symmetric $\phi$ profile,
\beq
\phi'+\frac{2}{\Lambda^3•}\frac{(\phi')^2}{r•}=\frac{\kappa}{2\pi\mpl r•}.
\label{3gcs}
\eeq
One can immediately read off the Vainshtein radius from this equation \footnote{Note that this decoupling-limit expression for the Vainshtein radius coincides with the one derived in the full DGP model in \cite{Lue:2001gc}.},
\beq
r_*\sim \(\frac{\kappa}{\mpl \Lambda^3•}\)^{1/2}.
\eeq
Using perturbation theory, one can readily solve for $\phi$ well outside the Vainshtein radius
to obtain the leading behavior of the scalar profile,
\beq
\phi\simeq\frac{\kappa•}{2\pi \mpl•}\ln \(\frac{r}{r_s}\), \qquad r\gg r_*,
\eeq
where again $r_s$ is a cutoff scale the finite thickness of the string.   Well within the Vainshtein radius, the nonlinear term in \eqref{3gcs} dominates, leading to
\beq
\phi\simeq \(\frac{\kappa \Lambda^3}{4\pi\mpl•}\)^{1/2} r+const., \qquad r\ll r_*.
\label{uvs}
\eeq

In the dual theory on the other hand, we can proceed in complete analogy to the above analysis. The last of the system \eqref{syst1} yields the following identity for such profiles,
\beq
b=-\bar\lambda-\frac{1}{2}r\bar \lambda'.
\eeq
Plugging this into the second of these equations and integrating, one finds that the equation determining $\bar\lambda$ is given as follows,
\beq
-\bar\lambda'+(\Lambda)^{3/2}\sqrt{\bar\lambda+b}=\frac{\kappa}{•2\pi\mpl r}.
\eeq
One can now solve this equation perturbatively, but perturbation theory now works well \textit{within} 
the Vainshtein radius, 
\beq
\bar\lambda=-\frac{\kappa•}{2\pi \mpl•}\ln \(\frac{r}{r_s}\)+\Lambda^{3/2}\(\frac{\kappa}{4\pi\mpl•}\)^{1/2} r+const.,
\eeq
arriving at an expression for $\phi$ which is valid in the UV,
\beq
\phi=\bar\phi+\bar\lambda=\Lambda^{3/2}\(\frac{\kappa}{4\pi\mpl•}\)^{1/2} r+const.~
\eeq
This is in complete agreement with the expression \eqref{uvs}, obtained from the original formulation.

\section{Conclusions}

We have studied a few examples of classical dualization for Nambu-Goldstone and galileon theories, 
which allows for a  perturbative formulation of the regimes in which the original theory becomes 
classically non-perturbative. Quantum mechanically, 
such a formulation is only valid in special cases, where  certain symmetries make it possible for classical 
nonlinearities to be strong, while keeping quantum corrections under complete control. 

Among scalar field theories, galileons perhaps represent the most remarkable examples of this, due to a powerful 
non-renormalization theorem \cite{Luty:2003vm,Nicolis:2008in,Hinterbichler:2010xn} protecting the 
leading part of the Lagrangian from quantum corrections. 
Hence, our results are justified for these theories.  Moreover,  it should be possible to generalize 
our approach to the higher galileon terms. 

In addition to capturing  many features of  the DGP model, galileons have emerged as an essential ingredient in 
the recently formulated ghost-free massive gravity models \cite{deRham:2010ik,deRham:2010kj}. The decoupling limit of 
these theories 
represents a certain (scalar-tensor) extension of the galileon with a more general structure, which however 
retains all the nice properties of the galileons, such as the presence of the Vainshtein mechanism and the non-renormalization 
theorems (for studies of cosmological and spherically symmetric solutions in the decoupling limit and beyond  
in ghost-free massive gravity, see \cite{deRham:2010tw,Nieuwenhuizen:2011sq,Koyama:2011xz,Koyama:2011yg,Chkareuli:2011te,
Gruzinov:2011mm,Chamseddine:2011bu,Mohseni:2011vv,D'Amico:2011jj,Gumrukcuoglu:2011zh,Gumrukcuoglu:2011ew,Berezhiani:2011mt,
Mirbabayi:2011aa} and references therein). Our method  should have a straightforward algorithmic generalization 
to those more general  Lagrangians obtained in \cite{deRham:2010ik}, and should be useful
for the development of the analog of the Parametrized Post Newtonian  formalism in massive 
gravity \cite {deRham:2010kj} (or for generic modified gravity models with extra scalar fields, such as the recently proposed Fab Four theory, \cite{fabfour}) and for systematically determining the observational consequences of the Vainshtein mechanism \cite{Hui:2009kc,Hui:2012jb}.

The method presented above might as well be useful in  studies of 
the proposal of Ref. \cite{gia1} (see \cite{gia3} for  treatment of the 
scattering problem in this context for the theories considered above).

Finally, it remains to be seen if quantization of the classical duals considered in 
the present work can lead to duals of the full quantum theory. Given the non-analytic  
nature of the dual theories that we 
obtained, quantization seems to be a nontrivial task. We expect the non-covariant decomposition
of the auxiliary fields used in Sections 2 and 3 to be a good starting point for bookkeeping 
of the degrees of freedom. The quantization procedure, may or may not force us to introduce 
additional dynamical degrees of freedom at the scale $\Lambda$.

\vskip 1cm 

\noindent {\bf Acknowledgments:} We thank  Lasha Berezhiani and Giga Chkareuli for useful discussions.
GG is supported by NSF grant PHY-0758032.  
KH is supported in part by NSF grant PHY-0930521, and by Department of Energy grant DE-FG05-95ER40893-A020, 
and would like to thank the department of physics at NYU for its hospitality during which time this work 
was completed. D.P. is supported by the US Department of Energy under contract DOE-FG03-97ER40546.

\appendix

\section{More General Interactions\label{nappendix}}

In this appendix we study a more general derivative self-interaction which is an arbitrary power of the field $\phi$, in order to show that the dual formulation always has non-analytic self-interactions with fractional powers of the fields, and that this fractional power is always less than 2, i.e. less than that of a mass term.

Consider a (ghost-free) Goldstone - \stu - type theory with a shift symmetry and a $Z_2$ symmetry.  A general interaction term containing only one derivative per field is given by the following,
\beq
\L=-\frac{1}{2•}(\p\phi)^2-\frac{\((\p\phi)^2\)^n}{2 n \Lambda^{4n-4}}+\frac{1}{\mpl}\phi T.
\label{general}
\eeq

The theory can be equivalently rewritten by integrating in the vector field  $\psi_\mu$,
\beq
\L=-\frac{1}{2•}(\p\phi)^2-\psi^\mu \p_\mu\phi+\(1-\frac{1}{2n•}\)\Lambda^{\frac{4n-4}{2n-1}} \( (\psi_\mu)^2    \)^{\frac{n}{2n-1}}+\frac{1}{\mpl}\phi T. \label{general2}
\eeq  
Integrating out $\psi_\mu$ recovers (\ref{general}).  Note that the power of $\psi_\mu$ in the interaction term is always $<2$.

The equations of motion that follow from (\ref{general2}) can be reduced to a form similar to the system \eqref{eqns2}. Using the same decomposition of the vector field as before, $\psi_\mu\to(\chi~,\psi^T_i+\p_i\psi)$, one finds that $\chi=\psi^T_i=0$. One combination of the fields, $\sigma=\phi+\psi$, in the presence of a point-like external source has the usual $1/r$ profile at all distances, whereas $\psi$ has a crossover Vainshtein scale due to nonlinearities. Moreover, the form of the dual action suggests that at small distances, nonlinearities in $\psi$ should be subdominant, providing a small perturbation over the $1/r$ potential. This can be checked explicitly by perturbatively solving the static equation of motion for $\psi$,
\beq
-\vec\nabla\cdot\( \vec\nabla\psi +\Lambda^{\frac{4n-4}{2n-1}} {\vec\nabla\psi}\( (\vec\nabla\psi)^2 \)^{-\frac{n-1}{2n-1}}  \)=\frac{1}{•\mpl}T~.
\eeq
For a static point-like source $T=-M\delta^3(\bf{x})$, this reduces to
\beq
\psi'+  \Lambda^{\frac{4n-4}{2n-1}} (\psi')^{\frac{1}{2n-1}}=\frac{M}{4\pi \mpl r^2}.
\eeq
Expanding $\psi$ into contributions of different order in powers of the non-linear interaction, 
\beq
\psi=\psi_0+\psi_1+..., 
\eeq
with $\psi_0=-{M\over 4\pi \mpl r}$, one finds that the first perturbation should satisfy the following equation\footnote{Note that the fractional powers of $-M/\mpl r^2$ in this equation are well-defined for any $n$.}, 
\beq
\psi_1' =   - \Lambda^{\frac{4n-4}{2n-1}} \(\frac{M}{4\pi\mpl r^2}\)^{\frac{1}{2n-1}},
\eeq
which is solved by the following expression,
\beq
\psi_1=-\frac{(2n-1)}{ (2n-3) }  \Lambda^{\frac{4n-4}{2n-1}}\(\frac{M}{4\pi\mpl}\)^{\frac{1}{2n-1}}r^{\frac{2n-3}{2n-1}}.
\eeq
One can estimate the crossover distance $r_*$ for the $\psi$ profile as the one for which the perturbation theory breaks down, 
\beq
r_*\sim \(\frac{•M}{ M_P•}\)^{1/2}{1\over \Lambda}.
\eeq
This scale is of the same order for any $n$, up to an weakly $n$-dependent multiplicative factor of order one.
The expression for the physical field $\phi$ inside the Vainshtein radius is
\beq
\phi=\sigma-\psi=-\psi_1=\frac{(2n-1)}{ (2n-3) }  \Lambda^{\frac{4n-4}{2n-1}}\(\frac{M}{4\pi\mpl}\)^{\frac{1}{2n-1}}r^{\frac{2n-3}{2n-1}}.
\eeq

\end{document}